\documentclass[
aps, prl,%
twocolumn,%
superscriptaddress,%
showpacs,%
preprintnumbers,%
amsmath,%
amssymb%
]{revtex4}

\usepackage{graphicx}
\usepackage{multirow}
\usepackage{epsfig}
\usepackage{float}

\begin{document}

\preprint{Physical Review Letters \textit{in press}.}

\title{Origin of Discrepancies in Inelastic Electron Tunneling Spectra of
Molecular Junctions}

\author{Lam H. Yu}
\author{Christopher D. Zangmeister}
\author{James G. Kushmerick}
    \email{james.kushmerick@nist.gov}
    \affiliation{National Institute of Standards and Technology, Gaithersburg, MD 20899}

\date{\today}

\begin{abstract}
We report inelastic electron tunneling spectroscopy (IETS) of
multilayer molecular junctions with and without incorporated metal
nano-particles.  The incorporation of metal nanoparticles into our
devices leads to enhanced IET intensity and a modified line-shape
for some vibrational modes.  The enhancement and line-shape
modification are both the result of a low lying hybrid metal
nanoparticle-molecule electronic level. These observations explain
the apparent discrepancy between earlier IETS measurements of alkane
thiolate junctions by Kushmerick \emph{et al.} [Nano Lett.
\textbf{4}, 639 (2004)] and Wang \emph{et al.} [Nano Lett.
\textbf{4}, 643 (2004)]. 
\end{abstract}

 \maketitle

Inelastic electron tunneling spectroscopy (IETS) is unique in that
it provides an \emph{in situ} vibrational probe of a molecular
electronic junction.  The IET spectrum represents a \emph{molecular
signature} for a junction, simultaneously proving that the molecule
of interest is present and giving insight into how the charge
carriers interact with the molecule\cite{Small_2_172}. Experimental
IET spectra can be analyzed with the aid of recently developed
computational techniques\cite{PRB_72_033408, PRL_95_146803,
JChemPhys_124_094704} to yield information on the nature of the
metal-molecule interfaces\cite{NatMat_5_901}, the orientation of the
molecules\cite{Nano_6_1693, Nano_6_1784}, and the position of the
molecules relative to the metallic
electrodes\cite{JPhysChemB_109_8519}.  With further refinements to
the interpretive modeling tools, IETS has the potential to become a
standard characterization technique for all molecular electronic
devices. However, recent IETS measurements of alkenethiolates in two
different test beds, crossed-wire\cite{Nano_4_639} and
nanopore\cite{Nano_4_643} junctions, have yielded markedly different
spectra.  The most obvious differences between these spectra are
their different lineshapes and the presence or absence of the C-H
stretching modes. While it was originally proposed that the
discrepancy in observing the C-H stretching modes resulted from
differences in the two molecules studied, namely a monothiol in the
crossed-wire experiment and a dithiol in the nanopore experiments,
it has subsequently been demonstrated both
experimentally\cite{Nano_6_2515} and theoretically\cite{Nano_5_1551}
that this difference alone is not sufficient to account for the
observed discrepancies.

In this letter, we report IET spectra of multilayer junctions with
and without incorporated metal nanoparticles (Figure 1).  The
incorporation of metal nanoparticles ($<$ 1 nm diameter) with low
lying electronic states into a large energy gap ($>$ 2 eV) molecular
layer can lead to what we coin enhanced IETS. The experimental
manifestations of enhanced IETS include orders of magnitude increase
in the IET spectral intensity and drastic changes in the lineshapes
of select vibrational modes.

\begin{figure}
    \includegraphics[width=50mm]{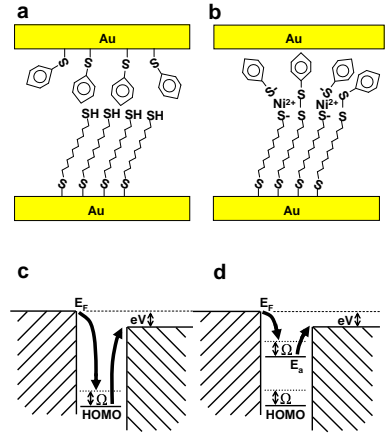}
    \caption{\small (Color online) Structural illustration of (a) the control device, where a monolayer of \textbf{BT} and \textbf{C10} is
    self-assembled onto either electrode, and (b) the molecule-particle-molecule
    device, where layers of \textbf{C10} and Ni and \textbf{BT} are sequentially self-assembled onto one electrode.  Energy diagram for
    inelastic electron tunneling via (c) a molecular orbital of the control device, and (d) a hybrid metal particle-molecule electronic
    resonance level, $E_a$, in the molecule-particle-molecule
    device. ($E_F$ is the Fermi energy, $\Omega$ is the energy of a vibration mode, and eV is the
    applied bias.) }
    \label{fig:Scheme}
\end{figure}

Fabrication of molecule-particle-molecule (MPM) multilayer
devices\footnote{MPM multilayer thin films are constructed by first
immersing a Au wire into a 3 mM ethanol solution of
1,10-decanedithiol for $\approx$14 h.  The wire is sonicated in
fresh ethanol for 5 min to remove any physisorbed molecules, then
placed in a 35 mM Ni(ClO$_{4}$)2*6H$_{2}$O ethanol solution for 3 h.
Following the thiol reaction with the Ni and another 5 min of
sonication, the wire is exposed to a 3 mM solution of benzenethiol
in ethanol for 14 h then sonicated for 5 min.} is based on previous
work done on Cu-dithiol\cite{Langmuir_13_5602, JPhysChemB_107_11721}
and Ni-dithiol\cite{Nano_6_2515} multilayer thin films. Cartoons
illustrating the structures of the MPM multilayer and the control
device consisting of a benzenthiol (\textbf{BT}) monolayer on one
electrode and a 1,10-decanedithiol (\textbf{C10}) monolayer on the
other electrode (without the incorporation of metal particles) are
shown in Fig.~\ref{fig:Scheme}. Details of the crossed-wire tunnel
junction apparatus used in our experiments have been previously
reported\cite{PRL_89_086802, Nano_4_639, Nano_6_2515}. The
crossed-wire apparatus is housed in a stainless steel vacuum chamber
that is evacuated and purged with He gas before being submerged into
a liquid He storage dewar.  All transport measurements are made
after the crossed-wire junction has been immersed in liquid He for
at least 1 h. Transport measurements are performed with standard ac
modulation techniques with two lock-in amplifiers that record the
first and second harmonics signals (proportional to $dI/dV$ and
$d^{2}I/dV^{2}$, respectively) simultaneously with the I-V
characteristics.  The amplitude of the ac modulation used is 4 mV
RMS.  To remove any dependence on the junction area, the spectra are
presented as the normalized amplitude $(d^{2}I/dV^{2})/(dI/dV)$.

\begin{figure}
    \includegraphics[width=75mm]{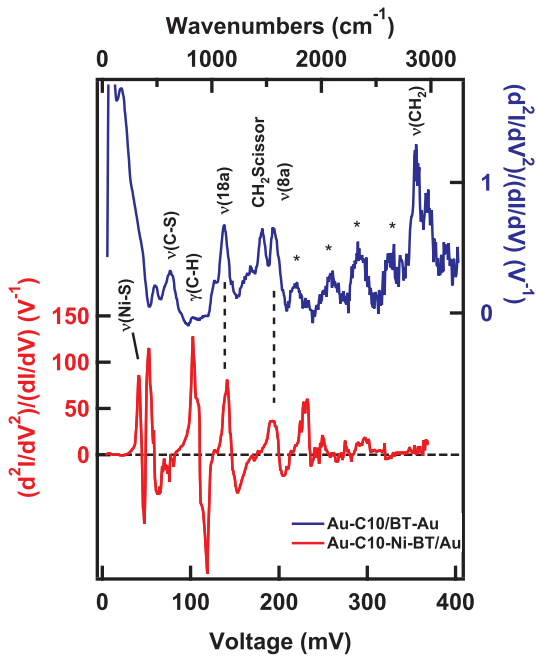}
    \caption{\small (Color online) IET spectra of a \textbf{C10}-Ni-\textbf{BT} multilayer
    junction (bottom) and a
    \textbf{C10}/\textbf{BT} control junction (top).  Mode assignments are based on
    previous experimental results
    and theoretical calculations.  Note $\upsilon$ = stretch,  $\gamma$ = out of plane
    bend and the phenyl modes are given in terms of Wilson-Varsanyi
    terminology.
    The asterisks mark modes likely associated with the free thiol group  of
    the \textbf{C10} layer.
    Modes associated with the metal nanoparticles and the \textbf{BT} layer dominate in the
    \textbf{C10}-Ni-\textbf{BT} junction.}
    \label{fig:IETS}
\end{figure}

Figure~\ref{fig:IETS} shows the IET spectra of a MPM junction and
that of a control junction. Since these junctions contain both
aliphatic and aromatic molecules, the observed spectra exhibit
vibrational modes associated with both moieties. We assign the
observed features to specific molecular vibrations by comparison to
previous IETS results\cite{Nano_4_639,Nano_6_2515,NatMat_5_901} and
density functional theory calculations. Although there are a number
of similarities between the spectra of the two junctions, there are
clear differences as well. The normalized amplitude of the MPM
junction IETS features are enhanced by over two orders of magnitude
compared with those of the control junction. In addition to the
enhancement there is a clear change in the peak shape. While all the
observed spectral features of the control junction have a simple
peak shape, the vibrational features in the MPM junction have
distinctive peak-derivative-like lineshapes. Such intensity
enhancement and peak-derivative lineshapes are never observed in IET
measurements on non-metal embedded crossed-wire
junctions\cite{Nano_4_639,Nano_6_2515}, nor in magnetic bead
junctions\cite{NatMat_5_901}. Another striking difference between
the two spectra in Fig.~\ref{fig:IETS} is that the aliphatic C-H
stretching mode near 360 mV is missing in the MPM junction spectrum.
Similarly, the zero-bias feature (ZBF)\footnote{The origin of the
ZBF has been discussed previously\cite{Nano_4_639,Nano_6_2515} and
is believed to originate from phonon interactions in the gold
wires.} which extends out to $\approx$ 50 mV in the control junction
is almost imperceptible in the MPM junction spectrum.  A change of
scale (not shown) reveals that the MPM junction does in fact contain
a ZBF that extends out to $\approx$ 30 mV with a normalized
amplitude of $\approx$ 1.5 V$^{-1}$, comparable in absolute
magnitude to that of the control device ZBF.

Figure~\ref{fig:Scheme}c and \ref{fig:Scheme}d show schematically
how the embedment of metal particles in a molecular junction can
give rise to a metal particle-molecule hybrid resonant energy level
close to the Fermi level, which could leads to the enhancement of
the IET signal. In the control device, the molecular vibrational
modes are coupled to the molecular orbitals of the constituent
molecules, but in the MPM junctions the molecular vibrational modes
are also coupled with a new hybrid level which lies much closer to
the Fermi level than the highest occupied molecular orbital (HOMO)
of the constituent molecular layers. With a resonant level close to
the Fermi level, higher order elastic cotunneling processes can
contribute substantially to the electronic transport near the onset
of an inelastic channel\cite{PRL_59_339, JVacSciTechA_6_331,
PRB_68_205406, Nano_4_1605}.  In a low-order perturbative treatment
of this effect, Baratoff and Persson found that a resonant level
close to the Fermi level could lead to larger IETS intensity and
substantial modification of the spectral lineshapes
\cite{PRL_59_339, JVacSciTechA_6_331}. According to this
perturbative model and more advanced theoretical treatments based on
the non-equilibrium Green's function (NEGF)
formalism\cite{Nano_4_1605, JChemPhys_121_11965}, the manner in
which the intensity and lineshapes of the IET spectrum of our MPM
junctions will be modified depends on the energy, $E_a$, and width,
$\Gamma$, of the metal particle-molecule hybrid resonance, and on
its coupling to the molecular vibration, $\delta E$. The
electron-phonon coupling term, $\delta E$, describes the shift of
$E_a$ under the influence of the excitation of a particular
vibrational mode.

Baratoff and Persson find that the relative change in conductance,
 $d\sigma/\sigma$, upon reaching the threshold for excitation of a molecular
 vibration, $|eV| = \Omega$, is proportional to
\begin{equation}\label{Eq1}
\frac{{\delta E ^2 }}{{\left( {E _a  - \overline{E _F}} \right)^2 +
\left( {\Gamma/2} \right)^2 }}\frac{{\left( {\overline{E _F} +
\Omega - E _a } \right)^2  - \left( {\Gamma/2} \right)^2 }}{{\left(
{\overline{E_F} + \Omega  - E _a } \right)^2 + \left( {\Gamma/2}
\right)^2 }} ,
\end{equation}
where $\overline{E_F} = E_F - eV$, with $E_F$ being the Fermi level
of the unbiased electrodes, $e$ is the charge of an electron, $V$ is
the applied bias and $\Omega$ is the energy of the vibrational
mode\cite{PRL_59_339, JVacSciTechA_6_331}. They show that, depending
on the position and width of the metal particle induced electronic
level, the normalized amplitude of the IETS feature can be very
large, and the lineshape can be a peak, a peak-derivative, or a dip.
The numerical modeling results of Galperin \emph{et al.} clearly
demonstrate the evolution of a IETS feature from a peak to a
peak-derivative to a dip as the position of the electronic resonance
level shifts (see Fig. 1 of Ref.~\onlinecite{Nano_4_1605}).

\begin{figure}
  \includegraphics[width=60mm]{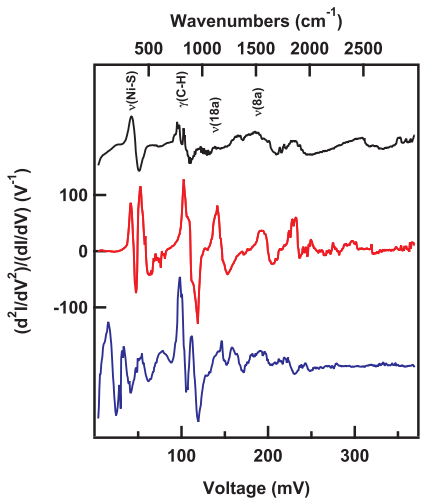}
  \caption{\small (Color online) IET spectra of three different \textbf{C10}-Ni-\textbf{BT}
  multilayer junctions (spectra are offset vertically for clarity).  The mode assignments
  are the same as those in Figure~\ref{fig:IETS}}
  \label{fig:Variability}
\end{figure}

An alternative mechanism that could give rise to the modified IET
spectra observed in the MPM junctions is resonant tunneling through
vibronic states of the molecular layer.  The signature of such a
process is equally spaced conductance peaks (or peak-derivative
features in the $d^2I/dV^2$ spectrum), where the energy spacing
between the features corresponds to a fundamental vibrational mode
of molecular layer\cite{JPhysChemB_102_1833, PRL_92_206102}.  Unlike
the IETS process we described above which involves virtual
excitation of vibronic states, this resonant process involves a real
transition to vibronic states.  While we cannot rule out resonant
vibronic excitation as the origin of the observed modified features,
there are two characteristics of our data which make such
interpretation difficult to justify.  First, we note the
peak-derivative features are not evenly spaced.  An analysis of the
distribution of energy spacings between features in our MPM data
reveals that successive features are separated from one another by
one of three different energy spacings roughly, 17 mV, 33 mV and 56
mV.  Furthermore, the onset of features in our MPM junction occurs
at bias $<$ 40 mV. For the resonant tunneling interpretation to be
correct the metal level would need to be practically degenerate with
the Fermi level of the electrodes\footnote{Although we do not
attribute the overall spectra to a Frank-Condon progression expected
for the resonant tunneling mechanism it is possible that the strong
feature at $\approx$100 mV is a combination of the second harmonic
of the Ni-S vibration and the phenyl C-H out of plane bending
mode\cite{JChemPhys_121_11965}.}.

Figure~\ref{fig:Variability} shows the IET spectra of three
different MPM junctions (The middle trace is the same as the MPM
spectrum shown in Fig.~\ref{fig:IETS}). While the same spectral
features dominate each spectrum, there is substantial variation in
the normalized amplitude and the lineshapes between the three
spectra. The bottom two spectra are $\approx$ 3.5 times more intense
than the top spectrum.  In light of Eq.1, such device to device
variations of IET spectrum intensity is not surprising since $E_a$,
$\Gamma$, and $\delta E$ depend strongly on the specific nanoscale
environment of each device.

\begin{figure}
  \includegraphics[width=70mm]{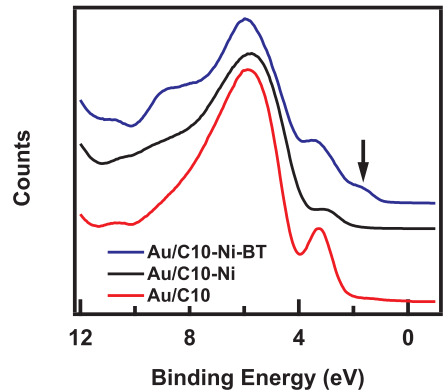}
  \caption{\small (Color online) Ultraviolet photoelectron spectra of Au/\textbf{C10}, Au/\textbf{C10}-Ni, and
  Au/\textbf{C10}-Ni-\textbf{BT} (spectra are offset vertically for clarity).  The arrow points to the hybrid electronic resonance
   arising from
  the metal coordination in the Au/\textbf{C10}-Ni-\textbf{BT} monolayer. Binding energies are referenced as positive for occupied states below
  the Fermi level (binding energy=0 eV). UP spectra were obtained with excitation via a He I line source (21.2 eV) with a hemispherical
  electrostatic analyzer (resolution $\approx$ 100 meV).}
  \label{fig:UPS}
\end{figure}

The presence of a low lying electronic resonance in our MPM
junctions is experimentally confirmed by ultraviolet photoelectron
spectroscopy (UPS). Figure~\ref{fig:UPS} shows the UP spectra of a
Au/\textbf{C10} (bottom), a Au/\textbf{C10}-Ni (middle) , and a
Au/\textbf{C10}-Ni-\textbf{BT} (top) molecular layer. In UPS,
one-photon photoelectron processes are employed to provide direct
measurement of the energies and characteristics of the occupied
electronic levels of the assembled molecular layer of interest. The
most intense peak on all three spectra at $\approx$ 5.9 eV is
attributed to the sigma bonds of the alkane
backbone\cite{JPhysChemB_107_11690}. The peak at 3.2 eV in the
Au/\textbf{C10} molecular film is attributed to the localized 3p
lone-pair electrons on the top-terminal S of the \textbf{C10}
molecules\cite{JPhysChemB_107_11690}. In the Au/\textbf{C10}-Ni film
this feature is attenuated and shifted to a lower binding energy of
$\approx$ 3 eV, consistent with the coordination of the Ni atoms to
the terminal S. The shoulder feature at $\approx$ 3.3 eV on the
Au/\textbf{C10}-Ni-\textbf{BT} spectrum is actually a convolution of
the localized 3p S lone-pairs and the HOMO of the benzenethiol
molecule\cite{Chris_unpublished}.

In contrast, the HOMO of the Au/\textbf{C10}-Ni-\textbf{BT}
molecular layer is characterized by the shoulder structure centered
at $\approx$ 1.5 eV.  The UP spectrum of a control molecular layer
composed of Au/\textbf{C10}-Ni-\textbf{C6}(not shown), where
\textbf{C6} represents 1-hexanethiol, shows no energy level near 1.5
eV.  This strongly suggests that this low lying electronic state is
a delocalized Ni-\textbf{BT} hybrid energy level.  This
interpretation is further supported by the observation that the Ni
and \textbf{BT} vibrational modes are the only modes \emph{enhanced}
in the \textbf{C10}-Ni-\textbf{BT} junction (Figs.~\ref{fig:IETS}
and \ref{fig:Variability}).

From the UPS data we extract $E_a$ for the MPM junctions as 1 eV
based on the onset energy of the hybrid energy level. An inspection
of the IETS literature suggests that $\delta$E ranges from 0.3 eV to
0.4 eV and $\Gamma$ ranges from 0.8 eV to  1
eV\cite{PRL_59_339,PRB_68_205406, JChemPhys_121_11965}.  Using these
parameters and Eq. 1, a rough estimate of d$\sigma$/$\sigma$ for the
$\nu$(8a) stretching mode at $\approx$ 185 meV of the MPM junctions
is 2$\%$ to 9$\%$. The experimental value of d$\sigma$/$\sigma$ for
this vibrational mode as extracted from the spectra shown in Fig. 3
is (7 $\pm$ 3) $\%$. For comparison the experimental value of
d$\sigma$/$\sigma$ for this same vibraional mode in the control
junction is 0.2 $\%$. The agreement between the estimated and
experimental $d\sigma/\sigma$ strongly suggests that the essential
physics of resonantly enhanced IETS is captured by Eq. 1 and the
formalism from which it is derived.

Based on the data presented here we feel that it is likely that the
nanopore devices of Wang \emph{et al.} contained nanoscale metal
islands within the 1,8-octanedithiol monolayer\cite{Nano_4_643}.
Incorporation of metal islands -- most likely formed during the top
contact deposition\cite{APL_84_4008} -- would explain the lineshapes
they observed as well as the lack of C-H stretching mode. More
recent IET spectra of nanopore devices with ferromagnetic electrodes
and 1,8-octanedithiol molecules\cite{APL_89_153105} again showed
characteristics consistent with metal nanoparticle incorporation.

In summary, we observed that the incorporation of metal
nanoparticles into molecular junctions can result in substantial
modification to the intensity and lineshape of the IET spectrum.  By
comparing the IET spectra of our MPM crossed-wire junctions with
those from nanopore devices, we conclude that the embedment of metal
nanoparticles into the molecular layer of the nanopore devices is
the likely cause of the differences previously observed in
crossed-wire\cite{Nano_4_639} and nanopore\cite{Nano_4_643}
junctions of similar molecular systems.

\begin{acknowledgments}
The authors thank L. Richter for useful discussions.  Financial
support from the DARPA MoleApps program is gratefully acknowledged.
\end{acknowledgments}

\bibliographystyle{apsrev}
\bibliography{Enhance}

\end{document}